\documentstyle[amsbsy,amstex,aps,epsf,multicol]{revtex}

\def\eqref#1{Eq.~(\ref{#1})}

\def\phi{\varphi}

\def\({\left(}
\def\){\right)}
\def\[{\left[}
\def\]{\right]}
\def\<{\left\langle}
\def\>{\right\rangle}
\def\<{\left\langle}
\def\>{\right\rangle}

\def\bea{\begin{eqnarray}}
\def\eea{\end{eqnarray}}

\makeatletter

\def\vlp{\mathopen{\boldsymbol{(}}}    
\def\vrp{\mathclose{\boldsymbol{)}}}   

\def\8{\infty}

\def\undertext#1{\vtop{\hbox{#1}\kern 1pt \hrule}}

\def\be{\begin{equation}}
\def\ee{\end{equation}}
\def\bea{\begin{eqnarray} & &}
\def\eea{\end{eqnarray}}

\title{A Solvable Model for Nonlinear Mean Field Dynamo }
\author{Stanislav~Boldyrev\thanks{E-mail: boldyrev@itp.ucsb.edu}
\\
{\em Institute for Theoretical Physics, Santa Barbara, California 
93106}}

\date{May 21, 2001}

\begin{document}

\input psfig.sty



\maketitle

\begin{abstract}
We formulate  a solvable model that describes generation and
saturation of mean magnetic field in a dynamo with kinetic helicity,
in the limit of large magnetic Prandtl number. 
This model is based on the assumption that the stochastic part of the
velocity field is Gaussian and white in time (the Kazantsev-Kraichnan
ensemble), while the regular part 
describing the back reaction of the magnetic field is chosen from
balancing the viscous and Lorentz stresses in the MHD Navier-Stokes
equation. The model provides an analytical explanation for previously
obtained numerical results.

\vspace{5mm} \noindent
\end{abstract}
\vspace{1mm}
\begin{multicols}{2}


{\bf 1}. 
Turbulent motion of a conducting fluid generates magnetic fields in
planets, stars, and interstellar medium. 
Observational~\cite{Lynden-Bell} 
and
theoretical~\cite{Parker,Kraichnan,Kazantsev,Kulsrud,Cowley,Chertkov} 
data
suggest that simultaneously with developing 
intense small-scale fluctuations, magnetic fields appear to be ordered
on scales much larger than the correlation length of the
turbulent velocity field. Explanation of the large-scale field
has remained 
a challenge for the astrophysical dynamo theory for a long time. A leading approach
suggests that the large-scale field is generated due to broken parity
in the turbulent velocity fluctuations. Mathematically, this means
that the fluctuating velocity field~$u$ possesses nonzero helicity, 
i.e., $\int u\cdot (\nabla\times u) \neq 0$.  
The analytical confirmation of the large-scale field generation 
is based on the splitting of the
magnetic field into the large scale and small scale components
(compared to the velocity correlation length), and
averaging the induction equation over the small scales~\cite{Moffatt}. 
This procedure leads to
the equation for the mean magnetic field~$\langle B^i \rangle$, 
which can be written in the general form:
\begin{eqnarray}
\partial_t \langle B^i \rangle=
\nabla \times (\alpha \, \langle B^i \rangle) + 
\beta \Delta \langle B^i \rangle + \dots
\label{alpha_beta}
\end{eqnarray}
Functions~$\alpha$ and~$\beta$ should be obtained from the
theory. As an estimate, one can accept~$\alpha \sim u$ and $\beta \sim
u\,l$, where $l$~is the characteristic scale of the velocity field. 
It is assumed that the mean field is changing slowly in 
 space, therefore the higher order space derivatives 
of~$\langle B^i \rangle$ are omitted in~(\ref{alpha_beta}). 
The crucial fact is that 
the coefficient~$\alpha$ describing the amplification of the mean
field is zero if
the velocity fluctuations possess no helicity~\cite{Moffatt}.

There is no general theory that would allow one to derive  
coefficients $\alpha$~and~$\beta$ from the microscopic equations. 
A self-consistent 
derivation of the mean field equation can be carried out under 
certain simplifying assumptions. One of them is the assumption 
about the geometry of the problem (velocity field is
assumed to be two-dimensional), which leads to a solvable model
introduced by Vainshtein~\cite{Vainshtein}.
Another assumption is about smallness of magnetic fluctuations 
compared to the mean 
field~(see, e.g.,~\cite{Biskamp,Gruzinov_Diamond}). 
However, analytical
and numerical results on the initial (kinematic) stage of the dynamo 
demonstrate that it is the small scale magnetic fluctuations that are
most energetic, not the large-scale mean
field~\cite{Kulsrud,Blackman}. Therefore, an analytical treatment 
of this opposite limit in a general 3-dimensional case is in demand. 

Moreover, the previous analytical and numerical investigations raised
the question of the so-called catastrophic $\alpha$-quenching in 
the mean-field
dynamo. The mean magnetic field was shown to saturate when its 
energy reaches the equipartition value
divided by the magnetic Reynolds
number,~$R_m$~\cite{Gruzinov_Diamond,Vainshtein_Cattaneo}. For 
the critical discussion of the~$\alpha$-quenching 
see also~\cite{Brandenburg2}. 
The magnetic resistivity thus played an essential role in the 
saturation mechanism. Astrophysical applications, e.g.,  
galactic dynamos, provide a unique physical setting of extremely large
magnetic Prandtl numbers (ratio of fluid viscosity to magnetic 
diffusivity), 
reaching 
$10^{14}$--$10^{22}$,
where the back reaction of the growing magnetic field can come into
play {\em before} the magnetic field reaches the diffusive scales. The
definitive numerical investigation of this problem requires very 
high resolution 
of the energy containing fine scales and is presently without 
the reach. An analytical understanding of
this limit is therefore important.

In the present paper we introduce a solvable model describing the 
nonlinear mean field dynamo in the limit of a large magnetic 
Prandtl number. We do not resort to any artificial scale separation 
procedure or to the special geometry, and do not 
assume that the magnetic fluctuations are small.
Instead, we make simplifying assumptions about
the velocity field. We assume that the microscopic fluctuations 
of the velocity field are
homogeneous, Gaussian, and short-time correlated, which constitutes
the so-called Kazantsev-Kraichnan ensemble~\cite{Kazantsev,Kraichnan},
while the the regular, ``back reaction," part is chosen from balancing
the viscous stress and the magnetic field stress. 
We also assume that the random velocity field possesses helicity. 
We believe 
that this model can provide some insight into  understanding of 
the mean-field dynamo, and can be complimentary to the previously
developed analytical treatments.


We show that in our model the
fluctuations of the magnitude of the magnetic field are independent 
of the fluctuations of its direction. In the presence of helicity, the
probability distribution function~(PDF) of the direction of the 
magnetic
field becomes nonstationary and anisotropic in~$x$ space. However, the
probability distribution function of the magnitude of the magnetic 
field does not depend on~$x$. Although the anisotropic part of
the PDF decays in the course of time, 
the {\em magnitude} 
of the magnetic field increases. The interplay of these two 
factors leads to the generation of the non-decaying 
mean magnetic field,~$\langle B^i \rangle$. This exact mechanism 
of the mean magnetic field
generation is the first main result 
of the paper.  The mean magnetic field is shown
to be determined by the geometry of the problem and by
the boundary conditions; the fact previously emphasized
by Blackman and Field~\cite{Blackman_Field}.

When the Lorentz force becomes strong enough, one cannot 
neglect its back
reaction on the velocity field. If the velocity field helicity is
small, 
 this back reaction affects
only the probability density function of the magnetic 
field magnitude, not the PDF of the 
magnetic field direction. 
As an example, we present a model based on 
the balance of viscous stress and magnetic stress, that can be
solved exactly and clearly demonstrates this saturation: anisotropy of
the PDF decays as before but the second factor,
the magnetic field
{\em magnitude}, saturates and cannot support the mean
magnetic field anymore. The $\alpha$~term thus becomes reduced. This 
mechanism  
of saturation of the growing mean magnetic field is the second main 
result of the paper. 

In Section~2 we formulate the model and proceed with the detailed 
derivation of the mean field equation. In Section~3 we suggest a
simple closure describing dynamo saturation within the present model.
Section~4 discusses possible generalizations of the model and
its limitations.

{\bf 2}. 
Let us first neglect the Lorentz force and assume that the
fluctuating velocity 
field~$u^i(x,t)$ is random, homogeneous, Gaussian, and short-time 
correlated:
\begin{eqnarray}
\langle u^i(x,t)u^k(x't')\rangle=\delta(t-t')\kappa^{ik}(x-x'),
\label{correlator}
\end{eqnarray} 
where due to the lack of mirror invariance we
have~$\kappa^{ik}(x-x')\neq \kappa^{ik}(x'-x)$. Magnetic field obeys
the induction equation:
\begin{eqnarray}
\partial_t B^i +u^k B^i_k - B^k u^i_k =\eta \Delta B^i,
\label{induction}
\end{eqnarray}
where lower indices denote derivatives with respect to the
corresponding spatial
coordinates, and we assume summation over the repeated indices. 
In the case of large magnetic Prandtl numbers (which is
usually the case in astrophysical settings and is considered here), 
we neglect the magnetic diffusivity~$\eta$. Using  
equation~(\ref{induction}) as the Langevin equation for magnetic field
fluctuations, one can easily derive the Fokker-Plank equation for the
probability distribution function of the magnetic field measured
at some point~$x$, $P(B^i; x,t)$:
\begin{eqnarray}
\partial_t P &=& \kappa_0 \Delta P +\kappa_2 \, B^2 \frac{\partial^2 P}{\partial
(B^i)^2} -\frac{2}{d+1} \kappa_2\, B^i B^k \frac{\partial^2 P}{\partial B^i \partial
B^k} \nonumber \\
 &+& 2\,g\,\varepsilon^{ikl}B^i \frac{\partial}{\partial B^k}
\nabla_l P.
\label{fokker-planck}
\end{eqnarray}
The coefficients in this equation can be read off from the following
expansion of the velocity correlation function~(\ref{correlator}) in
small argument~$y\equiv x-x'$ for incompressible velocity field:
\begin{eqnarray}
\kappa^{ik}(y)=\kappa_0\delta^{ik}-\frac{\kappa_2}{2}
\left( y^2\delta^{ik}-
\frac{2\, y^iy^k}{d+1} \right)+g\,\varepsilon^{ikl}y^l +\dots,
\label{correlator_expansion}
\end{eqnarray}
where $d$~is the space dimension ($d=3$), and the $g$~term describes
kinetic helicity. Note that the value of~$g$ should obey the
realizability condition
\begin{eqnarray}
g^2 \leq \frac{5}{8}\kappa_0 \kappa_2,
\label{realizability}
\end{eqnarray}
that follows from the fact that 
$\langle (a {\bf u} + \nabla\times {\bf u})^2 \rangle \geq 0$ for
any~$a$.

For completeness, we would like to present here the main steps of the
derivation of Eq.~(\ref{fokker-planck}) from Eq.~(\ref{induction}).
First, let us introduce the so-called $Z$~function defined
as~$Z(\lambda, x,t)= \exp\vlp i\lambda^i B^i(x,t) \vrp $. 
Obviously,  the average of this function over the random velocity
field,~$u$, or, equivalently, over the resulting magnetic 
field distribution~$P(B;x,t)$,
gives the Fourier transform of the distribution~$P(B;x,t)$ with
respect to~$B^i$. Due to Eq.~(\ref{induction}) the~$Z$ function 
satisfied the following differential equation:
\begin{eqnarray}
\partial_t  Z  &=&-i \lambda^i  u^k B^i_k Z 
+ i \lambda^i  u^i_k B^k Z \nonumber \\
 &\equiv & -u^k \frac{\partial}{\partial
x^k} Z +  u^i_k \lambda^i \frac{\partial}{\partial \lambda^k} Z  . 
\label{Z_function}
\end{eqnarray}
Let us now average this equation with respect to the random field~$u$.
To do this we first formally iterate the Eq.~(\ref{Z_function}) once, getting:
\begin{eqnarray}
\partial_t Z = u^k(t)\frac{\partial}{\partial x^k} 
&{}&\int\limits^{t}_{-\infty}
\left[ u^l(t') 
\frac{\partial}{\partial
x^l} Z(t') - u^i_l(t') \lambda^i 
\frac{\partial}{\partial \lambda^l} Z(t')\right] \nonumber \\ 
 - u^i_k(t) \lambda^i \frac{\partial }{\partial \lambda^k}
\int\limits^{t}_{-\infty}&{}& \left[ u^l(t') \frac{\partial}{\partial
x^l} Z(t') 
-  u^l_j(t') \lambda^l \frac{\partial}{\partial \lambda^j} 
Z(t')\right],
\label{iteration}
\end{eqnarray}
where the integration is performed with respect to~$t'$ and we do not
write the dependence on~$x$ in the right hand side
of~(\ref{iteration}). The average over~$u$ can now be done 
{\em independently} of the $Z$~function, since the $Z$~function
depends on the velocities taken at earlier times, $t'<t$, and due to
causality, cannot depend on~$u(t)$. We thus average $u$'s and $Z$'s
independently in~(\ref{iteration}), and Fourier-transforming the
resulting equation with respect to~$\lambda$, arrive at 
Eq.~(\ref{fokker-planck}).
 
Now, introducing new variables, the magnitude  of the magnetic
field $B$ and its direction $n^i=B^i/B$ ($n^2=1$), we cast the
Eq.~(\ref{fokker-planck}) into the form:
\begin{eqnarray}
\partial_t P=\kappa_0 \Delta P + {\hat L}_B P +{\hat L}_n P,
\label{fokker-planck_1}
\end{eqnarray}
where 
\begin{eqnarray}
{\hat L}_B&=&\kappa_2 \frac{d-1}{d+1}
\frac{1}{B^{d-1}}\frac{\partial }{\partial B}
B^{d+1}\frac{\partial}{\partial B} \quad , \label{l}\\
{\hat L}_n&=& \kappa_2 \left(
\delta^{ik}-n^in^k\right)\frac{\partial^2}{\partial n^i \partial n^k}
-\kappa_2 (d-1)n^i\frac{\partial }{\partial n^i}\nonumber \\
&+& 2\,g\,\varepsilon^{ikl}n^i\frac{\partial }{\partial n^k}\nabla_l\,
.\label{n}
\end{eqnarray}
When differentiating with respect to~$n^i$, we assume that all
components of~$n^i$ are independent. This is a legitimate procedure
since one can look for the solution
in the form~$P={\tilde P} \delta(1-n^2)$, and observe that the 
$\delta$~function factors out, i.e.,  ${\hat L}_n \, \delta(1-n^2) f = 
\delta(1-n^2) {\hat L}_n f $ for an arbitrary function~$f$.

The statistics of magnetic field magnitude and magnetic field
direction are independent if they are initially independent. Let us
thus look for the solution of the Fokker-Planck
equation~(\ref{fokker-planck}) in the factorized form
\begin{eqnarray}
P=P_B(B,t)\,G(n,x,t),
\label{P}
\end{eqnarray}
where~$P_B$ is the function satisfying the equation~
\begin{eqnarray}
\partial_t P_B={\hat L}_B
P_B .
\label{P_0}
\end{eqnarray}
This equation can be solved exactly to give the well-known $\log$-normal distribution of~$B$ (see e.g. \cite{boldyrev}), 
but we do not need this solution for our present purposes. Let us
concentrate on the equation for the~$G$ function
\begin{eqnarray}
\partial_t G=\kappa_0 \Delta G + {\hat L}_n \, G\, .
\label{G}
\end{eqnarray}
If the kinetic helicity is nonzero, the last term in the
operator~(\ref{n}) leads to  coupling of~$n$ and~$x$. A rigorous
analysis would require solving the eigenvalue problem for the operator
in the right hand side of~(\ref{fokker-planck_1}), which we are 
going to do
elsewhere. For now, we can accept that the $G$~function
in~(\ref{P}) corresponds to the {\em largest} eigenvalue of the
operator in the right hand side of~(\ref{G}). 
The relevant structure of the solution can be simply 
understood in
the following way. Let us first assume that~$G$ is independent 
of~$x$. One can then formally expand~$G$ in powers of~$n^i$:
\begin{eqnarray}
G(n,t)=1+ S^i n^i+ Q^{ik} n^i n^k+D^{ikl} n^i n^k n^l+\dots ,
\label{expansion}
\end{eqnarray}
where~$S$, $Q$, and $D$ are functions of~$t$. Plugging this expansion
in the Fokker-Planck equation~(\ref{G}), we easily check that the
higher the order of the expansion coefficient, the faster it decays, 
and therefore for large times
we can safely truncate the expansion~(\ref{expansion}) after the first
relevant term, $G=1+S^i n^i$. Higher order terms can however be
important for consideration of higher moments of the magnetic 
field but are irrelevant for our consideration of the mean magnetic 
field~$\langle B^i \rangle$. 
This truncated solution, describing anisotropy of the magnetic 
field  
distribution,  behaves  as
\begin{eqnarray}
S \propto \exp[ -\kappa_2 (d-1) t]. 
\label{s_function}
\end{eqnarray}
This
suggests to look for the general solution of~(\ref{G}) in the form
\begin{eqnarray}
G(n,x,t)=1+{\bar B}^i(x,t)\, n^i \, \exp[ -\kappa_2 (d-1)t ].
\label{G(x)}
\end{eqnarray}
Equation~(\ref{G}) now reduces to (in three dimensions)
\begin{eqnarray}
\partial_t {\bar B}^i=\kappa_0 \Delta {\bar B}^i +2\, g\, 
\nabla \times {\bar B}^i,
\label{B_0}
\end{eqnarray}
and we recover the equation for the mean field~(\ref{alpha_beta}), with the
eddy diffusivity~$\beta=\kappa_0$, and~$\alpha=2 g$. 

To reveal 
the physical sense of the function~${\bar B}^i(x,t)$, let us 
calculate the
mean magnetic field~$\langle B^i \rangle $ using the
distribution~(\ref{P}). At first sight, it should decay
since the magnetic field distribution function becomes more
and more isotropic in the course of time, and the isotropic
distribution has no preferred direction for the magnetic field~$B^i$. 
However, the
amplitude of the magnetic field grows making such an average
nonvanishing.  
Indeed, multiplying the equation~(\ref{P_0})
by~$B^d$  (magnitude~$B$ times the volume factor~$B^{d-1}$) 
and integrating with respect to~$B$ we get the average of
the magnitude 
\begin{eqnarray}
\langle B \rangle =B_0 \exp [\kappa_2 (d-1) t ], 
\label{magnitude_B}
\end{eqnarray}
where we assumed~$\langle B\rangle
=B_0$ at the initial moment. Finally, averaging with
respect to angles with the aid of the distribution~$G$, we obtain 
$\langle B^i \rangle = {\bar B}^i(x,t)\, B_0/d$. Note that the exponential 
factors cancel out exactly
in~(\ref{s_function}) and~(\ref{magnitude_B}). 

To conclude this section, we would like to explain the meaning of 
the averaging procedure 
in the above formulae. Mathematically, we average over the
Gaussian ensemble~(\ref{correlator}). Physically, the averaging is
understood to be performed over the space. More precisely, we average
over the scales much larger than the
velocity correlation length, but much smaller than the scales at which
the function~${\bar B}^i(x,t)$ is changing. Since on these scales the
velocity field is not correlated, we are effectively averaging over
non-correlated regions of
space, which is equivalent to the ensemble averaging. This is a 
sensible
physical procedure, since, for example, the scale of the mean 
galactic magnetic
field is~$> 1 \, kpc$, while the velocities are correlated at the
scales~$\leq 100 \, pc$. 

{\bf 3}. 
Equation~(\ref{B_0}) shows that the growth of
the mean magnetic field is determined by the eigenvalues of
the operator on the right hand side of~(\ref{B_0}). The solution
depends on the magnitude of the
helicity of the velocity field, on the geometry of the problem, and on
the boundary conditions. Space Fourier transform of the
Eq.~(\ref{B_0}) shows that the eigenvalues of the r.h.s. operator are:
$\lambda_1=-\kappa_0 k^2, \lambda_{2,3}=-\kappa_0 k^2 \pm 2 g k$.
Choosing $g$~to be positive, we obtain that the largest 
growth rate 
is  equal
to~$\lambda_{\mbox{max}}=g^2/\kappa_0$ and is
achieved at~$k=g/\kappa_0$ if this~$k$ is allowed in the system. 
Obviously, this growth can not proceed 
for infinitely long time, since
eventually the growing magnetic field will be brought to 
equipartition with the velocity field. We therefore need to introduce
the back reaction of the magnetic field to the velocity field. We do
not know any rigorous way of including this reaction into the
Eq.~(\ref{fokker-planck}). A simple physical model can be
suggested in the case when the mean magnetic field is much smaller
than the magnetic fluctuations. This can happen when the
following two conditions are met:  

(1) 
The helicity of the velocity field is small, i.e.,~$\kappa_0/l, 
\kappa_2 l \gg g$, where $l$~is the correlation length of the
velocity field. In this case, the generated mean
magnetic field is concentrated on the scales much larger than the
velocity correlation length, and is growing with the rate much 
smaller than that of the fluctuations growth. 

(2) The initial mean magnetic field is 
smaller than the initial magnetic fluctuations. The realizability 
condition~(\ref{realizability})
then ensures that the mean magnetic field is much 
smaller than the magnetic fluctuations at all consecutive times.

We now show that
under these assumptions  the equipartition is first 
achieved {\em locally} between the magnetic fluctuations and 
the velocity field. In a Lagrangian frame, magnetic field is rotated and stretched by 
the fluctuating velocity
gradient matrix,~$u^i_k$:
\begin{eqnarray}
\frac{\mbox{d}}{\mbox{d}t }B^i = u^i_k B^k.
\label{lagrange}
\end{eqnarray}
To model the back reaction, we want to 
represent this fluctuating gradient as some random part,~${\tilde
u}^i_k$ plus  
some regular part (that does not include the random
variable~${\tilde u}^i_k$) which is proportional to the magnetic
fluctuations~$B^i$. Field~${\bar B}^i$, being much smaller than the
fluctuations, would give higher order terms in~$g/\kappa_2 l$,
$gl/\kappa_0$, and should
not be relevant at least at the beginning of the saturation. 
We have the following traceless tensors at our
disposal:~$\varepsilon^{ikl}B^l$, $B^i_k$,
and~$T^{ik} \equiv B^iB^k-\delta^{ik}B^2/d$, which are 
constructed from the
local magnetic field. The first two tensors  can not be physically
relevant, since they do not contribute to the energy exchange between
the magnetic and velocity fields, which due to~(\ref{lagrange}) is
proportional to~$\langle B^iB^ku^i_k \rangle $. As we show below, the
choice of~$T^{ik}$ leaves the operator~${\hat L}_n$ intact and changes
only the~${\hat L}_B$ part of the Fokker-Planck
equation~(\ref{fokker-planck}), (\ref{fokker-planck_1}). 
But before we proceed with the
derivation we would like to give a more sound physical
 motivation for choosing the regular part of~$u^i_k$
in the form of~$T^{ik}$, as suggested in~\cite{Boldyrev_Brandenburg}. 

Let us assume that the dynamo saturation starts when the viscous
stress~$\nu \Delta
u^i$ in the right hand side of the incompressible MHD
Navier-Stokes equation balances the dynamical
stress~
$$\nabla_k \left(B^iB^k-\frac{1}{2}\delta^{ik}B^2 \right)-
\nabla_i\, p.$$ 
This balance seems to be reasonable at the {\em onset} of the back 
reaction, since
it is the smallest fluid eddies that are less energetic and that are
affected first by the growing magnetic field.
Also, assuming that the  pressure ensures
the incompressibility of the velocity field ($u^i_k $ must be 
traceless) we are left with  
\begin{eqnarray}
u^i_k=-\frac{1}{\nu}\left( B^iB^k-\frac{1}{d}\delta^{ik}B^2 \right) +
{\tilde u}^i_k,
\label{closure}
\end{eqnarray}
where the random part~${\tilde u}^i_k$ obeys~(\ref{correlator}), and
the regular part is just~$\frac{1}{\nu} T^{ik}$. The
structure of the regular part of the velocity gradient shows that the
magnetic field is stretched along itself, but not rotated. It is
instructive to see the possible physical meaning of this ansatz. 
Numerical results of~\cite{Brandenburg_etal,Cattaneo,Cattaneo1} suggest 
that the growing 
magnetic field
becomes organized in filaments that occupy small fraction of space.
The length of the filaments is of the order of the size of 
energy containing
eddies. The magnetic field amplitude inside the filaments is large, 
and the saturation  with the velocity fluctuations occurs first 
inside these filaments.
With such saturation, small-scale velocity fluctuations cannot 
bend the filaments, since this would lead to a large
restoring Lorentz force. The reduction of curvature of the filaments
during the saturation process has indeed been observed numerically by
Brandenburg~{\em et al}~\cite{Brandenburg_etal}. 
We thus have to again exclude the local 
rotation 
(terms $\varepsilon^{ikl}B^l$ and $B^i_k$) and leave only the 
stretching along the magnetic field lines which is given 
by~(\ref{closure}).

As we have mentioned, the ansatz~(\ref{closure}) does not change
the~${\hat L}_n$ part of the Fokker-Planck equation. The~${\hat L}_B$
part changes and takes the form:
\begin{eqnarray}
\partial_t P_B = {\hat L}_B P_B
 +\frac{d-1}{\nu d}\frac{1}{B^{d-1}}\frac{\partial }{\partial B}
B^{d+2} P_B.
\label{f-p_closure}
\end{eqnarray}
The saturated PDF can be found analytically as a stationary 
solution of this equation, and turns out to be Gaussian,
\begin{eqnarray}
P_B (B)=A \exp \left( -\frac{2}{3\nu \kappa_2} B^2 \right),
\label{gaussian}
\end{eqnarray} 
where $A$ is the normalization constant and~$d=3$, as oppose to the
log-normal distribution at the initial, kinematic, stage. Comparison 
with direct 
numerical 
simulations~\cite{Boldyrev_Brandenburg} shows that this gives 
the qualitatively correct 
behavior of the magnetic field PDF -- the PDF indeed saturates at 
the Gaussian.\footnote{A. Schekochihin was able to
find the exact {\em
time-dependent} solution of Eq.~(\ref{f-p_closure}), which 
allowed him to observe the evolution of~$P_B(B,t)$ from the kinematic
log-normal to the saturated Gaussian form. (Private communication,
2001.)} It is curious that the closure~(\ref{closure}), 
suggested here for the onset of the back reaction, also  works for the fully
saturated regime. 
The Gaussian stationary distribution of the magnetic field was 
earlier obtained in numerical simulations by Cattaneo~\cite{Cattaneo1} 
for the {\em central} part of 
the distribution function, while the {\em tails} of the PDF 
seemed to be
non-Gaussian. We thus expect our local ansatz~(\ref{closure}) to be 
valid only for
the central part of the PDF; indeed, not being able to capture the 
spatial structure of the magnetic field it cannot account
for the intermittency effects described by the non-Gaussian tails of
the PDF. Note, that Gaussianity of the magnetic field distribution was
also assumed in an analytical model for the mean magnetic field by
Subramanian~\cite{Subramanian}, see also numerics
in~\cite{Brandenburg}.

The mechanism of the saturation of helical dynamo 
with small mean field is thus the following. The growing 
fluctuations of the magnetic field
magnitude, described by~$P_B(B,t)$, saturate {\em locally} with the 
velocity fluctuations, i.e., as described by the above
model~(\ref{closure})-(\ref{f-p_closure}). 
After that the isotropisation of the magnetic field 
distribution proceeds faster than the amplitude growth,
and suppresses further growth of~$\langle B^i \rangle $: exponential
decay of~(\ref{s_function}) is not compensated by~(\ref{magnitude_B})
anymore, and due to the realizability condition~(\ref{realizability}),
can not be compensated by the growth rate~$\lambda_{\mbox{max}}$
either. This provides a mechanism for the
observed reduction
of the $\alpha$~factor. Remarkably, our 
analytical result based on ansatz~(\ref{closure}) is also in
accord with the physical picture of dynamo saturation advocated by
Cattaneo~\cite{Cattaneo,Cattaneo1}: saturation occurs 
first locally in filaments that gradually
fill the whole space.

Finally, it is instructive to see at what level the mean 
magnetic field should saturate. Due to~(\ref{gaussian}), the 
saturated value of magnetic fluctuations energy is 
estimated as the energy value of the smallest turbulent eddies,~$W\sim \nu
\kappa_2 \sim E/Re^{1/2} $, where~$E$ is the total fluid energy and~$Re$ is the 
Reynolds number of the fluid turbulence. The saturation time 
can then be found from~(\ref{magnitude_B}) to 
be~$t_s \sim \log (W/W_0)/(4 \kappa_2)$. Assuming that the mean 
magnetic field grows with the maximal possible growth 
rate,~$\lambda_{\mbox{max}}=g^2/\kappa_0$, we obtain for the saturated
energy of the mean field,~${\bar W}$:
\begin{eqnarray}
{\bar W}\sim  {\bar W}_0 \left( W/W_0 \right)^{\gamma} ,
\end{eqnarray}
where~$\gamma=g^2/(2 \kappa_0 \kappa_2)$. Setting~$\bar W_0 \sim W_0$,
and letting~$\gamma$  have the maximal possible value allowed 
by the
realizability condition~(\ref{realizability}), $\gamma_m=5/16$, 
we estimate:
\begin{eqnarray}
{\bar W}\lesssim E \, \left( W_0/E \right)^{1-\gamma_m}
Re^{-\gamma_m /2}.
\end{eqnarray}
We would like to stress the strong dependence of the saturated mean
magnetic field on the value of the {\em initial} magnetic field, and rather weak
dependence on the fluid Reynolds number. Since
in our model the back reaction starts {\em before} the resistive
scales are reached, the answer is independent of the magnetic
diffusivity. 

{\bf 4}. In the present paper we developed a framework for
analyzing the generation and saturation of the mean magnetic field,
which is based on the Kazantsev-Kraichnan ensemble for the velocity
field. Supplemented by the physically motivated ansatz for the
magnetic field back reaction, such a model accounts for
the growing  mean magnetic field exactly. 

Let us now discuss possible limitations of the approach. 
The solution of the mean-field equation requires the specification of
the particular geometry and can be different for different boundary
conditions. Of course, the condition of homogeneity of the velocity
fluctuations could be incompatible with real boundary conditions, 
which would  force us to consider more general velocity correlators,
$\kappa(x+x'; x-x')$ and to solve the eigenvalue problem analogous 
to~(\ref{G}). Another generalization may be necessary 
for the case 
of arbitrary, not small, magnitude of the mean magnetic field. 
In this case, the ansatz for the regular part of the velocity 
field can include 
large~$ {\bar B}^i$ field as well. Also, the neglected
tensor~$B^i_k$, though not leading to the energy transfer, can
be important for the isotropisation of the PDF in this case. Another
possible generalization is to consider a compressible velocity field.
All these questions are for the future.


I am indebted to Eric Blackman, Axel Brandenburg, Vladimir Pariev, 
and Dmitri 
Uzdensky for important suggestions and comments on both the physics 
and the style of the paper, and to Lars Bildsten and
Alexander Schekochihin  for the discussion of the results.

\end{multicols}
\end {document}